\documentclass[12pt]{article}

\usepackage{graphicx}

\begin{document}
\begin{center}
{\Large\bf \boldmath Mechanics of Quark Exchange in High-Energy \\Hadron Reactions at Forward Angles} 

\vspace*{6mm}
{M.V. Bondarenco}\\      
{\small \it  NSC Kharkov Institute of Physics \& Technology,\\ 1
Academicheskaya St., Kharkov 61108, Ukraine}
\end{center}

\vspace*{6mm}

\begin{abstract}
The 2-quark back-angle scattering mechanism is shown to reproduce
main features of high-energy $hh$ flavor-exchange reactions.
Prospects for reduction of the reaction matrix element to a form of
convolution of hadron wave functions with the hard scattering kernel
are discussed. Wave function models suitable for convolution with
the kernel, which is singular at small $x$, and the emerging
form-factor types, are discussed.
\end{abstract}

\vspace*{6mm}

Reactions of one unit of flavor exchange between two hadrons
scattering through small angles at high energy can occur, to LO in
Mandelstam $s$, due to exchange of a single pair of quarks, which
enter into a hard collision and completely interchange their large
longitudinal momenta (scatter to $180^\circ$ in c.m.s.). With no
rapidity gap between the hadron remnants and their new comoving
quarks created, they should with high probability recombine into a
new pair of hadrons. In the binary reaction channel the differential
cross-section, indeed, exhibits a forward peak of typical
$p_\perp\sim300$ MeV width. The peak decreases with the energy
rather slowly, as $s^{-1\div-2}$ -- approximately as does the
Born-level 2-quark back angle scattering differential cross-section
($\propto s^{-2}$); the rest of the energy dependence must be
attributed to reggeization effects.

The Born diagram of the head-on relativistic quark collision is
closely similar to that of real photon scattering on a quark, owing
to the similarity between hard gluon and hard quark propagators on
the light cone. Yet, high energy hadron reactions are much
\emph{easier to measure} then hard forward real Compton scattering
is, given the coupling constant factor $\alpha_s^2$ in the
cross-section instead of $\alpha_{EM}^2$, and no radiative or
diffractive background involved. On the theory side, it brings
complications, since eikonal (gauge link, Wilson line) phases in
hadron remnant interactions emerge, which spoil factorization in
terms of GPDs. However, that may actually happen in photon-hadron
interactions as well, and the $A^+=0$ gauge condition does not help
\cite{Brodsky}. As those links are to be dealt with anyway, it would
be better to constrain them using cross-checks from different
processes.

Forward Compton scattering is, also, not the best playground for
studying hadron spin effects, since quark SSA for it cancel in the
sum of two LO diagrams. As for 2-quark scattering in flavor
exchange, there is only one LO diagram, so transverse polarization
may be open, provided soft contributions supply necessary phase
shifts. Generally, soft effects can stem from eikonal links, and
also from initial and final state wave functions sandwiching the
\emph{complex} hard scattering kernel. Both eikonal and
initial/final state contributions are proportional to the
interaction strength, but still, it is not excluded that some of
them may happen to be more significant. To make a sensible decision,
it is instructive first to develop qualitative understanding of the
dynamics.

\section{Finite Longitudinal Shift + LS-Coupling}

Consider the act of the head-on quark collision in the rest frame of
one of the hadrons. Instead of backscattering, then, we have braking
of a fast quark and knockout of a resting one. The mediating hard
gluon, being shed from a fast particle, must propagate along the
same direction. Its range (Ioffe time \cite{Belitsky}) appears to be
collision energy independent and thereby, despite high virtuality of
the gluon, finite: $l_z\sim\left(M_{targ}x_{targ}\right)^{-1}\sim
R_H$. So, the stop point of the incident quark on average is shifted
with respect to the centre of the well by a distance commensurable
with the hadron radius, and with the simultaneous finite transverse
momentum transfer in the head-on collision, the captured quark
acquires an orbital angular momentum in the normal direction.

Of course, if by definition the final state, as well as the initial
one, must have zero orbital angular momentum ($s$-state), the quark
would be forbidden to orbit. But relativistically, the single-quark
$s$-state is the one having upper Dirac bispinor components
multiplied by spatial wave functions with $L=0$, but the lower ones,
by parity reasons, must have $L=1$, composing it with the spin
$S=\frac{1}{2}$ to yield the same total momentum $J=\frac{1}{2}$.
Now, to flip the spin of the $s$-state bound quark in a head-on
collision, it obviously suffices to change projection of $\bf L$ for
its lower Dirac bispinor components.

From the emerging spatial picture it seems clear that the effect is,
basically, \emph{bulk}, not due some fine local interference. So,
the implementation of real eikonal phases is unlikely to modify the
picture drastically. Opacity effects can be of consequence, but
their 3d distribution is rather uncertain. Therefore, a reasonable
strategy would be to begin with the neglect of eikonal factors, and
ultimately look for deriving constraints on them when a nonremovable
discrepancy with the data arises\footnote{Essentially, this is the
same attitude as in DGLAP, where, by the way, the role of eikonal
factors, to date, still has not become tangible, at typical $x$.}.
Thus we arrive at the representation of $hh$-collision amplitudes in
terms of wave function overlaps\footnote{Note that for elastic
scattering that is impossible in principle, because it is
\emph{caused} by soft exchanges. The minimum number of gluons
exchanged in the $t$-channel is two, and generally they need not be
all attached to the same parton line(though, conditions for the
opposite are sought in `heavy Pomeron' models).}.

A popular model of light-front WF with $LS$-interaction is that of
\cite{BrodskyDrell}. However, it is of perturbative origin, thence
inheriting support $x\in[0,1]$ and vanishing at the endpoints.
Therefore, no imaginary part can result in convolution of such wave
functions with the hard kernel $\frac{1}{x_1x_2+i0}$. An alternative
class of models which tend to give wave functions non-vanishing and
continuous at $x=0$ (with support in $x$ ranging, in principle, from
$-\infty$ to $+\infty$, and $x$ normalization being to the
single-constituent rest energy, not to that of the whole composite
system) assume non-interacting constituents moving in a
self-consistent field, static in their c.m.s.\footnote{At that, GPDs
near $x=0$ are not to be interpreted as quark densities at $x>0$ and
minus antiquark densities at $x<0$, at any rate not as those
extracted from DIS data using the parton model.} The static well is
unable to accommodate for recoil effects which must be crucial at
$-t>M^2$, but there is no experimental data in that far region
anyway. With wave functions continuous at $x=0$, the imaginary part
of the matrix element, in fact, logarithmically diverges in the hard
approximation, so it is not purely hard, but that introduces only a
weak dependence on the cutoff parameter.

The results obtained with a gaussian WF model were partially
presented at the conference, but prior to embarking at model
assumptions, it is desirable to exhaust all model-independent means.
To this end, one can admit that in general, $x$-dependence of $H$
and $E$ GPDs strongly differ\footnote{In contrast to the case of
$H$, $x$-dependence of $E$ is not constrained by DIS data. Note that
in some models \cite{Burkardt} those distributions are assumed to be
just proportional, but without any motivation beyond the similarity
of $t$-dependence of their $x$-integrals which are associated with
EM form-factors.}, and analyze consequences of factorization in
their terms.

\section{Factorization With Imaginary Contributions}

The factorization of the matrix element proceeds through
decomposition of the hard kernel
\[
\frac{1}{x_1x_2+i0}=\texttt{P}\frac{1}{x_1}\texttt{P}\frac{1}{x_2}-i\pi\frac{1}{|x_1|}\delta(x_2)-i\pi\delta(x_1)\frac{1}{|x_2|},
\]
and leads, for a representative reaction $np\to pn$, to the matrix
element
\begin{eqnarray*}
M_{fi}&\propto& w'^+_n\left(H_-(t)+\frac{i\sqrt{-t}}{2M}E_-(t)\sigma^N\right)w_pw'^+_p\left(H_-(t)+\frac{i\sqrt{-t}}{2M}E_-(t)\sigma^N\right)w_n\\
&-&iw'^+_n\left(H_+(t)+\frac{i\sqrt{-t}}{2M}E_+(t)\sigma^N\right)w_pw'^+_p\left(H_0(t)+\frac{i\sqrt{-t}}{2M}E_0(t)\sigma^N\right)w_n\\
&-&iw'^+_n\left(H_0(t)+\frac{i\sqrt{-t}}{2M}E_0(t)\sigma^N\right)w_pw'^+_p\left(H_+(t)+\frac{i\sqrt{-t}}{2M}E_+(t)\sigma^N\right)w_n\\
&+&\left(\tilde{H}_-^2(t)-2i\tilde{H}_+(t)\tilde{H}_0(t)\right)w'^+_n\sigma^Lw_pw'^+_p\sigma^Lw_n
\end{eqnarray*}
with
\[
H_-(t)=\texttt{P}\int\frac{dx}{x}H^u_p(x,0,t),
H_+(t,\lambda_x)=\int_{|x|>\lambda_x}
\frac{dx}{|x|}H^u_p(x,0,t)\approx H_+(t), H_0(t)=\pi H^u_p(0,0,t),
\]
and similarly for $E$ and $\tilde{H}$. Formally, this structure
resembles the sum of $t$-channel pole exchanges, but 2 of them
having \emph{imaginary} coupling constants. Polarization emerges due
to interference between the ``-" exchange and ``0+" and ``+0"
exchanges.

Note that $t$-dependences of ``-" form-factors dominated by typical
$x$ and of ``0" and ``+" form-factors fed by $x\approx0$ may be of
quite different width. Presence of several $t$-slopes is typically
observed in flavor exchange reactions, the pre-QCD interpretation
being the difference in masses of the exchanged mesons and onset of
central absorption \cite{Kane}. The quark backscattering theory,
however, does not contain correlated $q\bar{q}$ transverse
propagation.

Phenomenological consequences of $u$\emph{-channel} gluon
reggeization, and of possible contributions involving extra
$t$-channel color exchanges will be discussed elsewhere.

\textbf{Acknowledgement}. The author is grateful to organizers of
the conference for local hospitality during the meeting.



\begin{thebibliography}{99}\itemsep -1mm
\bibitem{Brodsky} S. J. Brodsky {\em et al}, PRD{\bf 65}, 114025 (2002).

\bibitem{Belitsky} A. V. Belitsky, and A. V. Radyushkin, Phys. Rep. {\bf 418}, 1 (2005).

\bibitem{BrodskyDrell} S. J. Brodsky, and S. D. Drell, PRD{\bf 22}, 2236 (1980).

\bibitem{Burkardt} M. Burkardt, Int. J. Mod. Phys. A{\bf 18} 173 (2003).

\bibitem{Kane} G. L. Kane, and A. Seidl, Rev. Mod. Phys.{\bf 48}, 309
(1976).

\end{thebibliography}
\end{document}